\documentstyle[aps,pre,multicol,amssymb,psfig]{revtex}

\begin{document}
\title{Coefficient of restitution of colliding viscoelastic spheres}
\author{Rosa Ram\'{\i}rez$^{1}$, Thorsten P\"oschel$^{2}$, Nikolai V.
  Brilliantov$^{2,3}$, and Thomas Schwager$^{2}$}

\address{$^1$Dpto. de Fisica. F.C.F.M., Universidad de Chile, 
  Casilla 487-3, Santiago, Chile}
\address{$^2$Humboldt-Universit\"at zu Berlin, Institut f\"ur Physik,
  Invalidenstr. 110, D-10115 Berlin, Germany\\ 
http://summa.physik.hu-berlin.de/$\sim$kies/}
\address{$^3$Moscow State University, Physics Department, Moscow
  119899, Russia}

\date{\today}
\maketitle
\begin{abstract}
  We perform a dimension analysis for colliding viscoelastic
  spheres to show that the coefficient of
  normal restitution $\epsilon$  depends on the impact velocity
  $g$ as $\epsilon=1-\gamma_1g^{1/5}+\gamma_2g^{2/5}\mp\dots$, in accordance
  with recent findings. We develop a simple theory to find explicit
  expressions for coefficients $\gamma_1$ and $\gamma_2$. Using these
  and few next expansion coefficients for $\epsilon(g)$ we construct a
  Pad\'e-approximation for this function which may be used for a wide
  range of impact velocities where the concept of the viscoelastic
  collision is valid. The obtained expression reproduces quite
  accurately the existing experimental dependence $\epsilon(g)$ for
  ice particles.
\end{abstract}


\begin{multicols}{2}

\section{Introduction}
The change of relative velocity of
inelastically colliding particles can be characterized by the
coefficient of restitution $\epsilon$. The normal component of the
relative velocity after a collision $g^{\prime}=
\vec{v}_{12}^{\,\prime}\cdot\vec{e}$ follows from that before the
collision $g=\vec{v}_{12}\cdot\vec{e}$ via

\begin{equation}
 g^{\prime}=-\epsilon g
\end{equation}
where $\vec{v}_{1}$, $\vec{v}_{2}$ and $\vec{v}_{1}^{\, \prime}$,
$\vec{v}_{2}^{\, \prime}$ are respectively the velocities before and
after the collision, while the unit vector $\vec{e} \equiv
\vec{r}_{12}/\left|\vec{r}_{12}\right|$ gives the direction of the
inter-particle vector $\vec{r}_{12}=\vec{r}_{1}-\vec{r}_{2}$ at the
instant of the collision.

\ From experiments as well as from theory it is well known that the
coefficient of normal restitution $\epsilon$ is not a constant but it
depends sensitively on the impact velocity
\cite{C1,C2,C3,C4,C5,C6,C7,C8,C9,C10,C11}. Although most of the
results in the field of granular gases have been derived neglecting
this dependence but using a velocity-independent coefficient of
restitution
(e.g.~\cite{Goldhirsch+Zanetti,Kadanoff1,Kadanoff2,Ernst1,Ernst2,Ernst3,Brey})
it has been shown that the impact-velocity dependence of the
coefficient of restitution has serious consequences for various
problems in granular gas dynamics
\cite{SaloLukkariHanninen:1988,HaemeenAnttilaLukkari:1980,SpahnSchwarzKurths:1997,PoeschelSchwager:1997b,BrilliantovPoeschel:1998d,SHB}.

The equation of motion for inelastically colliding 3D-spheres has been 
addressed in \cite{SHB,BSHP1,BSHP2}, where the Hertz contact law \cite{Hertz}
\begin{equation}
\label{Hertzlaw}
F_{\rm el} =\rho \, \xi^{3/2} \,,
\qquad \rho \equiv \frac{2Y}{3(1-\nu^2)}\sqrt{R^{\rm eff}}
\end{equation}
for the elastic inter-particle force has been extended to account for
the viscoelasticity of the material which causes the dissipative part
of the force
\begin{equation}
\label{Disslaw}
F_{\rm diss} =\frac32 A \, \rho \,\sqrt{\xi} \dot{\xi}\,.
\end{equation} 
Here,
$\xi$ is the compression of the particles during the collision
$\xi=R_1+R_2-\left|\vec{r}_1-\vec{r}_2\right|$ ($R_1$, $R_2$ and
$\vec{r}_1$, $\vec{r}_2$ are the radii and the positions of the
spheres), $Y$ and $\nu$ are respectively the Young modulus and the
Poisson ratio of the particle material, $R^{\rm eff} \equiv
R_1R_2/(R_1+R_2)$, and the dissipative parameter $A$ reads
\cite{BSHP1,BSHP2}

\begin{equation}
\label{A}
A=\frac13\, \frac{(3\eta_2-\eta_1)^2}{(3\eta_2+2\eta_1)}
\left[\frac{(1-\nu^2)(1-2\nu)}{Y \nu^2}\right]\,.
\end{equation} 
The viscous constants $\eta_1$, $\eta_2$ relate the dissipative stress
tensor to the deformation rate tensor \cite{BSHP1,BSHP2,LandauLifshits}.  The
same functional dependence of $F_{\rm diss}\left(\xi,\dot{\xi}\right)$ has been
obtained in \cite{KuwabaraKono,MorgadoOppenheim1,MorgadoOppenheim2} using a different
approach.

The equation of motion for inelastically colliding spheres reads,
 therefore, 
\begin{eqnarray}
  \label{eomotion}
&&  \ddot \xi +\frac{\rho}{m^{\rm eff}}\left( \xi^{3/2} +
\frac{3}{2}\,A\, \sqrt{\xi}\, 
\dot {\xi}\right)=0\,,\\
&&~~\mbox{with}~~~ \xi(0)= 0,~~\dot{\xi}(0)= g\,,\nonumber
\end{eqnarray}
and with $m^{\rm eff} \equiv m_1m_2/(m_1+m_2)$ ($m_1$, $m_2$ are the masses of
the colliding particles).  To obtain the dependence of the restitution
coefficient on the impact velocity for 3D-spheres the equation of
motion (\ref{eomotion}) was solved numerically \cite{C10,SHB,BSHP1,BSHP2} and
analytically~\cite{TomThor}, where the velocity-dependent restitution
coefficient has been obtained as a series in powers of $g^{1/5}$:
\begin{eqnarray}
  \label{epsilon}
  \epsilon=1&-&C_1 \left(\frac32\,A \right)
  \left(\frac{\rho}{m^{\rm eff}} \right)^{2/5}g^{1/5}\\
  &+& C_2\left(\frac32\,A \right)^2 \left(\frac{\rho}{m^{\rm eff}} 
  \right)^{4/5}g^{2/5} \mp \cdots\nonumber
\end{eqnarray}
The first coefficients $C_1=1.15344$ and $ C_2=0.79826 $ were
evaluated analytically and then confirmed by numerical simulations
\cite{TomThor}.

Although in \cite{TomThor} a general method of derivation of {\it all}
coefficients of the expansion (\ref{epsilon}) has been proposed, to
obtain these, extensive calculations have to be performed.  This
approach does not provide closed-form expressions for the
coefficients, but rather gives them in terms of convergent series
which are to be evaluated up to the desired precision.

In the present study we show that a dimension analysis allows
to obtain the functional form of the $\epsilon(g)$-dependence for the
elastic and dissipative forces.  Within the framework of this analysis
we reproduce
the dependence (\ref{epsilon}) up to numerical values of coefficients $C_k$.
A similar approach has been used by Tanaka et al.
\cite{Tanaka} to prove that the constant coefficient of restitution is
not consistent with physical reality (see also
\cite{C10,Taguchi:1992JDP}).  We also develop a simple approximative
theory, which gives a continuum fraction representation for
$\epsilon(g)$
and a closed-form expressions for $C_1$ and $ C_2$ with the same
numerical values as above. Using then coefficients $C_1, \ldots C_4$
(with $C_3$ and $C_4$ evaluated in Appendix in accordance with the general scheme
of Ref. \cite{TomThor}) we construct a Pad\'e-approximation, which
reproduces fairly well the experimental data for colliding
ice-particles \cite{C5}. 

\section{The dimensional analysis}
To perform the general dimensional analysis we adopt the following form
for the elastic and dissipative forces:
 
\begin{eqnarray}
F_{\rm el} &=& m^{\rm eff} D_1 \,\xi^{\alpha} \nonumber \\ 
F_{\rm diss} &=& m^{\rm eff} D_2\, \xi^{\gamma}\dot\xi^{\beta} \nonumber.
\end{eqnarray}
This general form (at least for small $\xi$ and $\dot{\xi}$) follows
from the fact that both elastic and dissipative forces vanish at
$\xi=0$ and $\dot{\xi}=0$, respectively. With these notations the
equation of motion for colliding particles reads
\begin{eqnarray}
&&  \ddot \xi +D_1\, \xi^{\alpha} + D_2\, \xi^{\gamma}\,  
\dot {\xi}^{\beta}=0\,,\\
&&\mbox{with}~~~ \xi(0)= 0,~~\dot{\xi}(0)= g\,,\nonumber
\end{eqnarray}
where $g$ has already been introduced. Now we choose as the
  characteristic length $\xi_0$ of the problem the maximal
compression for the elastic case. It may be found from the condition
that the initial kinetic energy $m^{\rm eff}g^2/2$ \cite{remark1} 
equals the maximal elastic energy $m^{\rm eff} D_1
{\xi_0}^{\alpha+1}/(\alpha+1)$, which yields
\begin{equation}
\xi_0 = \left
(\frac{\alpha+1}{2\, D_1} \right )^\frac{1}{1+\alpha}
{g}^\frac{2}{1+\alpha}\,, 
\end{equation}
Choosing then the { characteristic time} of the problem as 
$\tau_0=\xi_0/g$, we construct new dimensionless variables
\begin{equation}
  \label{dimlessvars}
  \hat{\xi} = \xi/\xi_0 \,, ~~~
  \dot{\hat{\xi}} = \dot\xi/g \,, ~~~
  \ddot{\hat{\xi}}= \frac{g^2}{\xi_0} \ddot\xi \, 
\end{equation}
and recast the equation of motion into dimensionless form:

\begin{eqnarray}
\label{dimles}
&&\ddot{\hat{\xi}} + \delta (g)\, \hat{\xi}^{\gamma}\,
\dot{\hat{\xi}}{}^{\beta} +\frac{1+\alpha}{2}\,\hat{\xi}^{\alpha} =0 ~~~
\mbox{with}\\
&&~~~ \hat{\xi}(0) = 0,~~\dot{\hat{\xi}}(0)=1  \nonumber \\
&&~~~ \hat{\xi}(\tau_c) = 0,~~\dot{\hat{\xi}}(\tau_c)=-\epsilon \nonumber 
\end{eqnarray}
In the last Eq. (\ref{dimles}) we supplemented the pre-collisional
initial conditions at $\tau=0$ with the after-collisional conditions
at $\tau=\tau_c$ ($\tau$ is the dimensionless time and $\tau_c$ is the 
dimensionless duration of the
collision). These  follow just from the definition of the restitution
coefficient.  We point out that all dependence on the initial
impact velocity on any quantity of the problem, including $\epsilon$
(this is just the dimensionless after-collisional velocity) comes only
through the constant $\delta$, which reads
\begin{equation}
\delta(g) = 
D_2\, \left (\frac{1+\alpha}{2\, D_1}\right )^\frac{1+\gamma}{1+\alpha}\,
g^{\frac{ 2 (\gamma -\alpha)}{1+\alpha}+\beta}\,.
\end{equation}
Hence, $\epsilon(g)=\epsilon \left( \delta(g) \right)$. 
A similar result for $\epsilon \to 0$, $\beta=1$ and $\alpha =3/2$ has 
been obtained in \cite{Falcon98}.

If we assume that the restitution coefficient does not depend on the 
impact velocity $g$ then follows

\begin{equation}
2\left(\gamma -\alpha\right)+\beta\,\left ( 1+\alpha\right ) = 0\,.
\label{condition}
\end{equation}
For a linear dependence of the dissipative force on the velocity, i.e. for 
$\beta=1$ (this seems to be the most realistic for small $\dot{\xi}$) 
one obtains a constant restitution coefficient for

\begin{itemize}
\item the linear elastic force, $F_{\rm el} \sim \xi$, 
i.e. $\alpha=1$. The condition (\ref{condition}) implies $\gamma=0$, and
thus the linear dissipative force $F_{\rm diss} \sim \dot{\xi}$.

\item the Hertz law for 3D-spheres (\ref{Hertzlaw}) 
$\alpha=3/2$, therefore $\gamma=\frac14$  and  
$F_{\rm diss} \sim  \dot\xi\, \xi^{1/4}$ provides a constant restitution 
coefficient. 
\end{itemize} 

We now ask the question: What kind of $\epsilon(g)$ dependence
corresponds to the forces which act during collisions of viscoelastic
particles?  It may be generally shown \cite{BSHP1,BSHP2,remark2} that the
relation

\begin{equation}
\label{eldiss}
F_{\rm diss}= A\, \dot\xi \frac{\partial}{\partial{\xi}} F_{\rm el}(\xi) \, 
\end{equation}
between the dissipative and elastic forces with the dissipative
constant $A$ given in eq.~(\ref{A}) holds, provided three conditions
are met~\cite{BrilPoshprep}:
\begin{enumerate}
\item[(i)] The elastic part of the stress tensor depends linearly on
  the deformation tensor~\cite{LandauLifshits}.
\item [(ii)] The dissipative part of the stress tensor depends
  linearly on the deformation rate tensor~\cite{LandauLifshits}.
\item[(iii)] The conditions of quasistatic motion are provided, i.e.
  $g \ll c$, $\tau_{\rm vis} \ll \tau_c$ \cite{BSHP1,BSHP2} (here $c$ is the
  speed of sound in the material of particles, $\tau_{\rm vis}$ is
  relaxation time of viscous processes in its bulk).
\end{enumerate}

\ From this follows that $\beta=1$, $\gamma=\alpha-1$, and thus 
the constant restitution coefficient may be observed only for 
collisions of cubic particles with surfaces normal 
to the direction of collision.  We wish to emphasize that this 
conclusion comes from the
general analysis of viscoelastic collisions.

Let us discuss now collisions between spheres with elastic and
dissipative forces as given by (\ref{Hertzlaw}) and (\ref{Disslaw}),
respectively. For these we have $m^{\rm eff}D_1= \rho$, $\alpha=3/2$
and $m^{\rm eff}D_2= \frac32 A \, \rho$, $\gamma=1/2$, $\beta=1$ which
yields the functional dependence for $\delta(g)$ and $\epsilon(g)$ respectively:

\begin{eqnarray}
  \label{eq:deltadef}
  \delta&=&\frac{3}{2}\left(\frac{5}{4}\right)^{3/5}A\left(\frac{\rho}{m^{\rm eff}}\right)^{2/5}g^{1/5}\\
  \epsilon&=&\epsilon \left(A \left(\frac{\rho}{m^{\rm eff}}\right)^{2/5} g^{1/5} \right)
\label{eq:epsilongeneral}
\end{eqnarray}
(skipping the prefactor of $\delta(g)$ in the last equation) in accordance with (\ref{epsilon}) as found previously. 

\section{The restitution coefficient for spheres}
Using $\frac{d}{dt}=\dot{\hat{\xi}}\frac{d}{d\hat{\xi}}$
it is convenient to write the equation of motion for a collision in the form
\begin{eqnarray}
&&\frac{d}{d \hat{\xi}} \left(\frac12 \dot{\hat{\xi}}^2 + 
\frac12 \hat{\xi}^{5/2} \right) =
-\delta \dot{\hat{\xi}} \hat{\xi}^{1/2}  
=\frac{dE(\hat{\xi})}{d\hat{\xi}} \nonumber \\
&& \hat{\xi}(0)=0; \qquad \dot{\hat{\xi}} (0) =1
\end{eqnarray}
where we introduce the mechanical energy
\begin{equation}
E \equiv \frac12 \dot{\hat{\xi}}^2 + \frac12 \hat{\xi}^{5/2}\,.
\end{equation}
To find the energy losses in the first stage of the collision, which
starts with zero compression and ends in the turning point with
maximal compression $\hat{\xi}_0$
\begin{equation}
\label{Elosdir}
\int_0^{\hat{\xi}_0} \frac{dE}{d\hat{\xi}} d\hat{\xi}= 
-\delta \int_0^{\hat{\xi}_0} \dot{\hat{\xi}} \hat{\xi}^{1/2} d\hat{\xi}
\end{equation}
one needs to know the dependence of the compression rate
$\dot{\hat{\xi}}$ as a function of the compression $\hat{\xi}$.

For the case of elastic collisions, the maximal compression is
$\hat{\xi}_0=1$, according to the definition of our dimensionless
variables. Hence, the dependence $\dot{\hat{\xi}} (\hat{\xi})$ follows
from the conservation of energy:
\begin{equation}
\dot{\hat{\xi}} (\hat{\xi})=\sqrt{1- \hat{\xi}^{5/2}} \, .
\end{equation}
The velocity $\dot{\hat{\xi}}$ vanishes at the turning point
$\hat{\xi}=1$.  For inelastic collisions the maximal compression
$\hat{\xi}_0$ is smaller than $1$, therefore, one can write { an
  approximation} relation for the inelastic case:
\begin{equation}
\dot{\hat{\xi}} (\hat{\xi}) \approx \sqrt{1- (\hat{\xi}/\hat{\xi}_0)^{5/2}}
\end{equation} 
which also gives vanishing velocity $\dot{\hat{\xi}}$ at the turning
point $\hat{\xi}_0$.  Integration in Eq. (\ref{Elosdir}) may be
performed yielding

\begin{equation}
\label{direct}
\frac12 \hat{\xi}_0^{\,5/2}-\frac12 = -\delta\, d\, \hat{\xi}_0^{\,3/2}
\end{equation} 
where we take into account that $E(\hat{\xi}_0)=\frac12
\hat{\xi}_0^{5/2}$, $E(0)=\frac12 \dot{\hat{\xi}}(0)=\frac12$ and
introduce a constant

\begin{equation}
  d \equiv \int_0^1 x^{1/2} \sqrt{1-x^{5/2}}=
  \frac{ \sqrt{\pi}\, \Gamma \left(\frac35 \right)}{5 
    \,\Gamma\left(\frac{21}{10} \right)}\,.
  \label{eq:defofd}
\end{equation} 
Consider now the inverse collision, which is defined as a collision
which starts with velocity $\epsilon\, g$ and ends with velocity $g$.  
According to the concept of the inverse collision
introduced in \cite{TomThor} (which is a useful auxiliary model), it is
characterized by a negative damping (the energy ``is pumped'' into the
system during the collision). The maximal compression $\hat{\xi}_0$ is
the same in both collisions, the direct and the inverse.

Rescaling equation of motion for the inverse collision in the very same way 
as for the direct collision yields

\begin{eqnarray}
&& \frac{dE(\hat{\xi})}{d\hat{\xi}} =
+\delta \dot{\hat{\xi}} \hat{\xi}^{1/2}  \nonumber \\
&& \hat{\xi}(0)=0\,, \qquad \dot{\hat{\xi}} (0) =\epsilon\,.
\end{eqnarray}
This suggests the following approximative relation for
$\dot{\hat{\xi}} (\hat{\xi})$ during the inverse collision:

\begin{equation}
\dot{\hat{\xi}} (\hat{\xi}) \approx \epsilon \,
\sqrt{1- (\hat{\xi}/\hat{\xi}_0)^{5/2}}\,,
\end{equation} 
with the additional prefactor $\epsilon$, which is the initial
velocity in the inverse collision.

Integration of the energy {\it gain} for the first stage of the
inverse collision (which equals up to its sign the energy loss in the
second stage of the direct collision \cite{TomThor}) may be performed
just in the same way as for the direct collision, yielding the result
\begin{equation}
\label{inverse}
\frac12 \hat{\xi}_0^{5/2}-\frac{\epsilon^2}{2} = +\epsilon\, \delta\, d\, \hat{\xi}_0^{3/2}\,,
\end{equation} 
where we again use $E(\hat{\xi}_0)=\frac12 \hat{\xi}_0^{\,5/2}$ and
$E(0)=\frac12 \epsilon^2$.  Multiplying Eq. (\ref{direct}) by
$\epsilon$ and summing it up with Eq. (\ref{inverse}) we obtain a
simple approximative relation between the restitution coefficient and
the (dimensionless) maximal compression:
\begin{equation}
\epsilon= \hat{\xi}_0^{\,5/2}  
\end{equation}
Substituting this into Eq. (\ref{direct}) we arrive at an equation 
for  the restitution coefficient
\begin{equation}
\label{eqforeps}
\epsilon+2\delta\, d\, \epsilon^{3/5}=1\,.
\end{equation} 
The formal solution to this equation may be written as a 
continuum fraction (which does not diverge in the limit $g\to
\infty$):
\begin{equation}
\label{eqforeps1}
\epsilon^{-1}=1+2\delta\, d\, 
\left( 1+ 2\delta d \left(1+ \cdots \right)^{2/5} \cdots \right)^{2/5}\,.
\end{equation} 

Another way of representation of the restitution coefficient $\epsilon$ is a series expansion in terms of $\delta$. For practical applications it is
convenient to return to dimensional units. We define the characteristic velocity $g^*$ such that 
\begin{equation}
\delta \equiv \frac{1}{2d} \left ( \frac{g}{g^*} \right )^{1/5}\,,
\end{equation}
with $d$ beeing defined in Eq. (\ref{eq:defofd}). Using the definition 
(\ref{eq:deltadef}) together with Eq. (\ref{eq:defofd}) we find for the 
characteristic velocity
\begin{equation}
  \left(g^{*}\right)^{-1/5}=\frac{\sqrt{\pi}}{2^{1/5}5^{2/5}}\frac{\Gamma\left(3/5\right)}{\Gamma\left(21/10\right)}\left(\frac{3}{2}A\right)\left(\frac{\rho}{m^{\rm eff}}\right)^{2/5}\,.
\end{equation}
Evaluating the numerical prefactor finally yields
\begin{equation}
  \left(g^{*}\right)^{-1/5}=1.15344\,\left(\frac{3}{2}A\right)\left(\frac{\rho}{m^{\rm eff}}\right)^{2/5}\,.
\end{equation}
Note that the numerical constant $1.15344$ has to be equal to $C_1$ in Eq. (\ref{epsilon}). 

With this new notation the restitution coefficient reads
\begin{eqnarray}
\label{eqforeps2x}
\epsilon&=&1-a_1\left(\frac{g}{g^*}\right)^\frac{1}{5}+a_2 
\left(\frac{g}{g^*}\right)^\frac{2}{5} \nonumber 
-a_3\left(\frac{g}{g^*}\right)^\frac{3}{5} \\&+& 
a_4 \left(\frac{g}{g^*}\right)^{4/5} + \cdots\,,
\end{eqnarray}
with  $a_1=1$, $a_2=3/5$ (which are exact values), $a_3=6/25=0.24$, $a_4=7/125 = 0.056, \cdots$ (which deviate from the correct ones, see below). Comparing (\ref{eqforeps2x}) with (\ref{epsilon}) we conclude that our simple approximative theory
reproduces exactly the coefficients $C_1$ and $C_2$, which were found before using
extensive analysis \cite{TomThor}.

We also performed rigorous but elaborated calculations according to the general scheme of \cite{TomThor} to find the exact coefficients (details are given in the Appendix)
\begin{equation}
  a_3=0.315119\,, ~~a_4=0.161167\,,
\end{equation}
or, respectively,
\begin{equation}
C_3 = -0.483582\,, ~~C_4=0.285279\, .
\end{equation}
Hence, we  observe that while the first two coefficients $a_1=1$ and $a_2=3/5$ are correctly obtained from the
approximative theory, the next approximated coefficients $a_3$, $a_4$ differ from the exact ones.

For practical applications such as Molecular Dynamics simulations,
however, the expansion (\ref{eqforeps2x}) is of limited value, due to
its divergence for high impact velocities, $g \to \infty $. According
to the velocity distribution function there is a certain probability
that the relative velocity $g$ of colliding particles exceeds the limit of
applicability of (\ref{eqforeps2x}). Therefore, we use the obtained
coefficients to construct a Pad\'e-approximation for $\epsilon(g)$
which reveals the correct limits of the boundary conditions,
$\epsilon(0)=1$ and $\epsilon(\infty)=0$. Since the dependence
$\epsilon(g)$ is expected to be a smooth, monotonically decreasing
function, we choose a ``1-4'' Pad\'e-approximation
\begin{equation}
\label{Padegen}
\epsilon
=\frac{1+d_1\left(\frac{g}{g^*}\right)^\frac15}{1+d_2
\left(\frac{g}{g^*}\right)^\frac15+d_3\left(\frac{g}{g^*}
\right)^\frac25+d_4\left(\frac{g}{g^*}\right)^\frac35+d_5
\left(\frac{g}{g^*}\right)^\frac45}\,.
\end{equation}
Standard analysis yields the coefficients $d_k$ in terms of
the coefficients $a_k$ \cite{Tabcom} (see Table~\ref{tab:tab})
\end{multicols}
\begin{table}[htbp]
    \leavevmode
    \begin{tabular}{ll}
      $d_0=a_4-2a_3-a_2^2+3a_2-1$ \\
      $d_1=\left[1-a_2+a_3-2a_4+(a_2-1)(3a_2-2a_3)\right]d_0^{-1}$& 
      $=2.5839$\\
      $d_2=\left[ (a_3-a_2)(1-2a_2)-a_4\right]d_0^{-1}$& $=3.5839$\\
      $d_3=\left[ a_3+a_2^2(a_2-1)-a_4(a_2+1)\right]d_0^{-1}$& 
      $=2.9839$\\
      $d_4=\left[ a_4(a_3-1)+(a_3-a_2)(a_2^2-2a_3)\right]d_0^{-1}$& 
      $=1.1487$\\
      $d_5= \left[ 2(a_3-a_2)(a_4-a_2a_3)-(a_4-a_2^2)^2-
a_3(a_3-a_2^2)\right]d_0^{-1}$& 
      $=0.3265$
    \end{tabular}
    \caption{The coefficients of the Pad\'e formula (\ref{Padegen}) 
as derived from the coefficients $a_k$ .}
    \label{tab:tab}
\end{table}
\begin{multicols}{2}
  Using the characteristic velocity $g^*=0.32 \, cm/s$ for ice as a
  fitting parameter we could reproduce fairly well the experimental
  dependence of the restitution coefficient of ice as a function of
  the impact velocity $g$ in the whole range of $g$
  (Fig.~\ref{fig:1}).  The discrepancy with the experimental data at
  small $g$ follows from the fact that the extrapolation expression,
  $\epsilon=0.32/g^{0.234}$ used in \cite{C5} has an unphysical
  divergence at $g \to 0$ and does not imply the fail of the theory
  for this region. The scattering of the experimental data presented
  in \cite{C5} is large for small impact velocity according to
  experimental complications, hence the fit formula of \cite{C5} cannot be
  expected to be accurate enough for too small velocities. Moreover in
  the region of very small velocity possibly other than viscoelastic
  interactions influence the collision behavior, e.g. adhesion.

  \begin{minipage}{8cm}
\begin{figure}[htbp]
\centerline{\psfig{figure=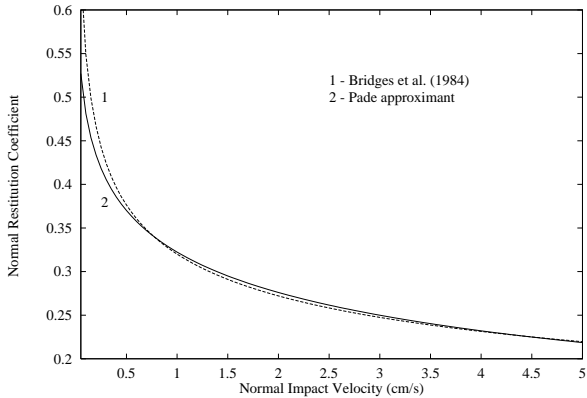,width=7.8cm,angle=270}}  
  \caption{Dependence of the normal restitution coefficient  on the 
    impact velocity for ice particles. The solid line -- experimental
    data of [5], 
    dashed line -- the Pad\'e-approximation (\ref{Padegen}) with the
    constants given in the Table and with the characteristic velocity
    for ice $g^*=0.32 \, cm/s$.}
  \label{fig:1}
\end{figure}
\end{minipage}

\section{Conclusion} 
We developed a dimensional analysis for the inelastic collision of
spherical particles. We could show that
for 3D-spheres the functional form for $\epsilon(g)$
agrees with that derived previously \cite{TomThor} using a much
more complicated approach.  Using a simple approximative theory we
found a continuum-fraction representation for $\epsilon(g)$ and obtained
explicit expressions for the coefficients of the series expansion of
the restitution coefficient in terms of the impact velocity.  The
first two coefficients in this series coincide with that found
previously by numerical evaluation. We report also a few next
coefficients which we have derived within the general approach of a
previous study \cite{TomThor}.  Using the first four coefficients of
this series expansion we constructed a Pad\'e-approximation for
$\epsilon(g)$. It reproduces fairly well the experimental data for
colliding ice particles.  The obtained relation for the restitution
coefficient may be used for a wide range of the impact velocities,
provided that the energy loss during a collision is attributed to
viscoelasticity and that all the other dissipative processes (plastic
deformation, fragmentation of particles) may be ignored.

\acknowledgments The authors want to thank W. Ebeling and L.
Schimansky-Geier for discussion. The work was supported by Deutsche
Forschungsgemeinschaft through grant Po 472/3-2 and by FONDECYT Chile,
through project 02960021.

\section{Appendix}

The general method of derivation of the expansion coefficients  
$C_k$ has been given in \cite{TomThor}. Here, we briefly sketch the main lines of derivation, 
and provide some details for the particular case of $C_3$ and $C_4$. 
Since the method of derivation is based on the collection of terms 
with different dependence on the initial velocity $g$, it is convenient
to use a scaling, somewhat different from that used before for the 
dimensional analysis. Namely, we rescale the time as
$t^{\prime} = (\rho /m_{\rm eff})^{2/5}g^{1/5}t$ and the 
length as $x =(\rho /m_{\rm eff})^{2/5} \xi$ to recast Eq. (\ref{eomotion})
into the form \cite{AttentionScales}
\begin{equation}
  \label{eq:eqnofmotionbeta}
  x^{\prime\prime}+\alpha g^{-1/5}x^{\prime}\sqrt{x}+g^{-2/5}x^{3/2}=0\,,
\end{equation}
with $\alpha \equiv \frac{3}{2}A (\rho /m_{\rm eff})^{2/5}$ and using all
the notations introduced previously. The initial conditions for the 
rescaled Eq. (\ref{eq:eqnofmotionbeta}) now read $x(0)=0$ and 
$x^{\prime}(0)=g^{4/5}$. For simplicity of notations we will keep 
in what follows $t$ for the rescaled time. As it was shown in \cite{TomThor},
the trajectory may be expanded in terms of $\sqrt{t}$ as
\begin{eqnarray}
&&x(t)= b_1 t^{1/2}+b_2t+b_3t^{3/2}+b_4t^{2} \\
&&~~~~~~~~+b_5t^{5/2}+b_6t^{3}+b_7t^{7/2}+\dots\,. \nonumber 
\end{eqnarray}
Clearly, both, $b_1$ and $b_3$ should be zero to avoid divergence of velocity 
and acceleration at $t=0$. At the same time  $b_2=g^{4/5}$ and $b_4=0$ due to the equation 
of motion at vanishing compression. This yields
\begin{equation}
\label{x(t)1}
x(t)= g^{4/5}t+b_5t^{5/2} +b_6t^{3}+b_7t^{7/2}+\dots\,.
\end{equation}
\ From (\ref{x(t)1}) one obtains $x^{\prime}(t)$ and $x^{\prime \prime}(t)$ 
which are to be substituted into the equation of motion 
(\ref{eq:eqnofmotionbeta}). One also needs $\sqrt{x}$ and $x^{3/2}$; the 
expansions for these in terms of $\sqrt{t}$ read
\begin{equation}
  \sqrt{x}=g^{2/5}t^{1/2}\!+\!\frac{b_5}{2g^{2/5}}t^{2}\!+\!
  \frac{b_6}{2g^{2/5}}t^{5/2}
  \!+\!\frac{b_7}{2g^{2/5}}t^{3}+\dots
\end{equation}
and  
\begin{equation}
  x^{3/2}=g^{6/5}t^{3/2}+ 
  \frac{3}{2}g^{2/5}b_5t^{3}+\frac{3}{2}
  g^{2/5}b_6 t^{7/2}+\dots\,.
\end{equation}
Inserting the expansions for $x^{\prime}(t)$, $x^{\prime \prime}(t)$, 
$\sqrt{x}$ and $x^{3/2}$ into (\ref{eq:eqnofmotionbeta}) and collecting 
the orders of $t$ we obtain
\begin{eqnarray}
0&=&\left(\frac{15}{4}b_5+\alpha g^{1/5}\right) t^{1/2}+6\,b_6t+
\left(\frac{35}{4}b_7+1\right)t^{3/2}
\nonumber\\ 
&+&\left(12\,b_8+3\alpha g^{1/5}b_5\right)t^{2}+\left(\frac{63}{4}b_9
+\frac{7}{2}\alpha g^{1/5}b_6\right)t^{5/2}\,.
\end{eqnarray}

This suggests the coefficients: 
\begin{eqnarray}
b_5&=&-\frac{4}{15}\alpha g^{1/5}\\
b_6&=&0\\
b_7&=&-\frac{4}{35}\\
b_8&=&\frac{1}{15}\alpha^2 g^{2/5}\\
b_9&=&0 \,,
\end{eqnarray}
so that the solution for the trajectory finally 
reads
\begin{eqnarray}
\label{trajectory}
x(t)&=& g^{4/5} t-\frac{4}{15}\alpha g t^{5/2}-\frac{4}{35} g^{4/5} t^{7/2}\nonumber\\
&+&\frac{1}{15}\alpha^2 g^{6/5}t^{4}+\dots \,.
\end{eqnarray}

In order to get the higher orders, which is conceptionally simple
but requires extensive calculus, we wrote a program~\cite{program},
which by formula manipulations performs exactly the steps we
described above and which is able to find the trajectory up to any
desired order. 

Generally, it is convenient to write the solution as a series
\begin{equation}
x(t)=g^{4/5} \left ( x_0(t)+\alpha g^{1/5} x_1(t)+\alpha^2 g^{2/5}x_2(t)+\dots\, \right ).
\label{sugested}
\end{equation}

Here $x_0(t)$ is a ``zero-order'' trajectory, which refers to the case of
undamped collision, the  ``first-order'' trajectory, $x_1(t)$, accounts for 
damping in linear (with respect to $\alpha$) approximation, 
the ``second-order'' trajectory, $x_2(t)$, corresponds to the next 
approximation $\sim \alpha^2$, etc. Here we give our results for  
these ``$n$-order'' trajectories up to $n=3$, obtained using the above 
mentioned program up to the order $t^{11}$:

\begin{eqnarray}
\label{x1x2x3}
&&x_0=t-\frac{4}{35}t^{7/2}+\frac{1}{175}t^6-\frac{22}{104125}t^{17/2}+\frac{52}{8017625}t^{11}\nonumber\\
&&x_1=-\frac{4}{15}t^{5/2}+\frac{3}{70}t^5-\frac{713}{238875}t^{15/2}+\frac{61216}{42639187}t^{10}\nonumber\\
&&x_2=\frac{1}{15}t^4-\frac{937}{75075}t^{13/2}+\frac{871}{808500}t^9\\
&&x_3=-\frac{38}{2475}t^{11/2}+\frac{43943}{13513500}t^8-\frac{1184627}{3594591000}t^{21/2}\nonumber
\end{eqnarray}

To proceed we need to find the maximal compression 
$x_{\rm max}$, which is reached at time $t_{\rm max}$. The point of maximal 
compression is a turning point of the trajectory, where the velocity is
zero. Therefore the condition 

\begin{equation}
\label{xmax}
x_{\rm max}^{\prime}(t_{\rm max})=0
\end{equation}
holds at this point. With the above expression for the trajectory 
Eqs. (\ref{sugested},\ref{x1x2x3}), the last Eq. (\ref{xmax}) 
is an equation to determine $t_{\rm max}$, which may be 
then used to find $x_{\rm max}$. This equation, however is a high-order 
algebraic equation for $\sqrt{t_{\rm max}}$, which is not generally 
solvable. On the other hand, for the undamped collision $t_{\rm max}$
equals one-half of the collision duration $t_c^{0}$ and both 
quantities of interest are known \cite{LandauLifshits}: 
\begin{eqnarray}
&&t_{\rm max}^{0}=\frac{t_c^{0}}{2}=
\left(\frac{4}{5}\right)^{3/5}
\frac{\Gamma\left(\frac{2}{5}\right)\Gamma\left(\frac{1}{2}\right)}{2\,\Gamma\left(\frac{9}{10}\right)} \nonumber\\ 
&&x_0\left(\frac{t_c^0}{2}\right)=\left(\frac{5}{4}\right)^{2/5}
\end{eqnarray}

For a viscoelastic collision $t_{\rm max}$ certainly differs from ${t_c^{0}}/{2}$, 
so that $t_{\rm max}={t_c^{0}}/{2} + \delta t$. If the dissipation parameter 
$\alpha$ is not large, the deviation $\delta t$ is presumably small, therefore we expand
$x^{\prime}(t_{\rm max})=x^{\prime}\left(\frac{t_c^{0}}{2} + \delta t \right)$ 
in terms of $\delta t$:
\begin{eqnarray}
\label{xprdelta}
&&g^{-4/5}\,x^{\prime}(t_{\rm max})= \\
&&=\left[x_0^{\prime}\left(\frac{t_c^0}{2}\right)+\delta tx_0^{\prime\prime}\left(\frac{t_c^0}{2}\right)+\frac{\delta t^2}{2}x_0^{\prime\prime\prime}\left(\frac{t_c^0}{2}\right)+\dots
 \right]\nonumber\\
&&+\alpha g^{1/5}\left[x_1^{\prime}\left(\frac{t_c^0}{2}\right)+\delta tx_1^{\prime\prime}\left(\frac{t_c^0}{2}\right)+\frac{\delta t^2}{2}x_1^{\prime\prime\prime}\left(\frac{t_c^0}{2}\right)+\dots\right]\nonumber\\
  &&+\alpha^2 g^{2/5}\left[x_2^{\prime}\left(\frac{t_c^0}{2}\right)+\delta tx_2^{\prime\prime}\left(\frac{t_c^0}{2}\right)+\dots\right]\nonumber\\
  &&+\alpha^3 g^{3/5}\left[x_3^{\prime}\left(\frac{t_c^0}{2}\right)+\dots\right]+\dots
=0\,,
\nonumber
\end{eqnarray}
where we use representation (\ref{sugested}) for the trajectory. The deviation $\delta t$, 
vanishes at $\alpha=0$ and, thus, suggests the expansion in terms of $\alpha$:

\begin{equation}
  \label{eq:ansatztmax}
\delta t=\tau_1\alpha+\tau_2\alpha^2+\tau_3\alpha^3 \dots
\end{equation}
Substituting $\delta t$, given by Eq. (\ref{eq:ansatztmax}), 
into (\ref{xprdelta}) and collecting 
terms of the same order of $\alpha$ yields
\begin{equation}
 Y_0+\alpha Y_1+\alpha^2 Y_2 +\alpha^3 Y_3 +\cdots =0\,,
\end{equation}
with the abbreviations
\end{multicols}
\begin{eqnarray}
  \label{xviatau}
  &&Y_0= x_0^{\prime}\left(\frac{t_c^0}{2}\right) \\
  &&Y_1=\tau_1x_0^{\prime\prime}\left(\frac{t_c^0}{2}\right)+g^{1/5}x_1^{\prime}\left(\frac{t_c^0}{2}\right) \nonumber\\
  &&Y_2= \tau_2x_0^{\prime\prime}\left(\frac{t_c^0}{2}\right)+\frac{\tau_1^2}{2}x_0^{\prime\prime\prime}\left(\frac{t_c^0}{2}\right) 
+g^{1/5}\tau_1x_1^{\prime\prime}\left(\frac{t_c^0}{2}\right)+g^{2/5}x_2^{\prime}\left(\frac{t_c^0}{2}\right) \nonumber\\
  &&Y_3=\tau_3x_0^{\prime\prime}\left(\frac{t_c^0}{2}\right)+\tau_1\tau_2x_0^{\prime\prime\prime}\left(\frac{t_c^0}{2}\right)+\frac{\tau_1^3}{6}x_0^{\prime\prime\prime\prime}\left(\frac{t_c^0}{2}\right)
+g^{1/5}\tau_2x_1^{\prime\prime}\left(\frac{t_c^0}{2}\right)+g^{1/5}\frac{\tau_1^2}{2}x_1^{\prime\prime\prime}\left(\frac{t_c^0}{2}\right)
+g^{2/5}\tau_1x_2^{\prime\prime}\left(\frac{t_c^{0'}}{2}\right)+g^{3/5}x_3^{\prime}\left(\frac{t_c^{0'}}{2}\right)\,. \nonumber
\end{eqnarray}
The conditions $Y_k=0$ for $k=0,\ldots 3$ together with Eq. (\ref{xviatau})
allows to express the constants $\tau_1$, $\tau_2$, $\tau_3$, etc., in terms 
of functions $x_1(t)$, $x_2(t)$, $x_3(t)$, etc., and their time derivatives 
taken at time $\left({t_c^{0}}/{2}\right)$:

\begin{eqnarray}
\label{tau1tau2}
&&  \tau_1=-g^{1/5}\frac{x_1^{\prime}\left(\frac{t_c^0}{2}\right)}{x_0^{\prime\prime}\left(\frac{t_c^0}{2}\right)}\\
&&  \tau_2=g^{2/5}\left[-\frac{x_1^{\prime 2}\left(\frac{t_c^0}{2}\right)x_0^{\prime\prime\prime}\left(\frac{t_c^0}{2}\right)}{2\, x_0^{\prime\prime 3}\left(\frac{t_c^0}{2}\right)}+\frac{x_1^{\prime}\left(\frac{t_c^0}{2}\right)x_1^{\prime\prime}\left(\frac{t_c^0}{2}\right)}{x_0^{\prime\prime 2}\left(\frac{t_c^0}{2}\right)} 
-\frac{x_2^{\prime}\left(\frac{t_c^0}{2}\right)}{x_0^{\prime\prime}\left(\frac{t_c^0}{2}\right)}\right] \nonumber
\end{eqnarray}
We do not write the expression for $\tau_3$, since due to the special properties of the problem, i.e.
due to the fact that $x_0^{\prime}\left({t_c^0}/{2}\right)=0$ the value $\tau_3$ is not needed 
for calculation of $\epsilon$ up to fourth order of $\alpha$. The functions 
$x_1(t)$, $x_2(t)$, $x_3(t)$ are known and given by Eqs. (\ref{x1x2x3}), so 
that the constants $\tau_1$ and  $\tau_2$ may be found explicitly.

Writing the maximal compression as 
\begin{eqnarray}
&&x_{\rm max}=g^{4/5} \left [ x_0\left(\frac{t_c^0}{2}+\delta t\right)+
\alpha g^{1/5}x_1\left(\frac{t_c^0}{2}+\delta t\right)
+\alpha g^{2/5}x_2\left(\frac{t_c^0}{2}+\delta t\right)+
\alpha^3g^{3/5}x_3\left(\frac{t_c^0}{2}+\delta t\right) \, \right ] , 
\end{eqnarray}
and expanding this in terms of $\delta t$, using then representation of 
$\delta t$ as $\delta t= \alpha \tau_1 +\alpha^2 \tau_2 + \cdots$, 
with $\tau_1$, $\tau_2$ from (\ref{tau1tau2}) and collecting terms 
of the same order of $\alpha$ we obtain
\begin{equation}
\label{xmaxdir1}
x_{\rm max} = g^{4/5}\left ( y_0+\alpha g^{1/5}y_1+\alpha^2 g^{2/5}y_2+\alpha^3 g^{3/5}y_3 \right )\,,
\end{equation}
where $y_0, \ldots y_3$ are pure numbers:
\begin{eqnarray}
&&y_0=x_0\left(\frac{t_c^0}{2}\right) = 1.093362 \\
&&y_1=x_1\left(\frac{t_c^0}{2}\right) = -0.504455  \\
&&y_2=\left[x_2\left(\frac{t_c^0}{2}\right)-\frac{1}{2}\frac{x_1^{\prime 2}
\left(\frac{t_c^0}{2}\right)}{x_0^{\prime\prime}\left(\frac{t_c^0}{2}\right)}\right]
=0.260542 \\
&&y_3=\left[x_3\left(\frac{t_c^0}{2}\right)-\frac{x_1^{\prime}\left(\frac{t_c^0}{2}\right)x_2^{\prime}\left(\frac{t_c^0}{2}\right)}{x_0^{\prime\prime}\left(\frac{t_c^0}{2}\right)}
+\frac{1}{2}\frac{x_1^{\prime 2}\left(\frac{t_c^0}{2}\right)x_1^{\prime\prime}\left(\frac{t_c^0}{2}\right)}{x_0^{\prime\prime}\left(\frac{t_c^0}{2}\right)}-\frac{x_1^{\prime 3}\left(\frac{t_c^0}{2}\right)x_0^{\prime\prime\prime}}{x_0^{\prime\prime 3}\left({t_c^0}/{2}\right)}\right]
=-0.136769
\end{eqnarray}
and where we use expressions (\ref{x1x2x3}) for $x_1(t)$, 
$x_2(t)$ and $x_3(t)$.

To calculate the coefficient of restitution one has to use the concept of inverse collision, as was introduced in \cite{TomThor} and discussed in previous chapters of the present study. One obtains the solution of this inverse collision by replacing $g\to \epsilon g$ for the initial velocity and $\alpha\to-\alpha$ for the 
dissipative coefficient. In particular, this applies to the maximal compression
of the inverse collision 
$x_{\rm max}^{\rm inv}=x_{\rm max}(g\to \epsilon g, \alpha\to-\alpha)$. 
 For consistency 
one has to require the maximum compressions for direct and inverse collision 
to be equal, i.e.
\begin{equation}
x_{\rm max}^{\rm inv}=x_{\rm max}\,,
\end{equation}
or using (\ref{xmaxdir1}),
\begin{eqnarray}
\label{genforeps}
&&\epsilon^\frac{4}{5} g^\frac{4}{5} \left ( y_0-\alpha \epsilon^\frac{1}{5} g^\frac{1}{5} y_1+
\alpha^2 \epsilon^\frac{2}{5}g^\frac{2}{5}y_2-\alpha^3 \epsilon^\frac{3}{5}g^\frac{3}{5}y_3
+\cdots \right )
= g^\frac{4}{5} \left (y_0+\alpha g^\frac{1}{5}y_1+\alpha^2 g^\frac{2}{5}y_2+\alpha^3 g^\frac{3}{5}y_3 +\cdots \right )\,.
\end{eqnarray}

Eq. (\ref{genforeps}) is in fact an algebraic equation for 
$\epsilon^{1/5}$, which may not be generally solved. For this reason 
we write $\epsilon$ as an expansion of $\alpha g^{1/5}$, which is the only combination in which both parameters appear
\begin{eqnarray}
\label{eq:epsilonansatz}
&&\epsilon=1+C_1\alpha g^{1/5}+C_2\left(\alpha g^{1/5}\right)^2 
+C_3\left(\alpha g^{1/5}\right)^3+C_4\left(\alpha g^{1/5}\right)^4+\dots
\end{eqnarray}
and substitute (\ref{eq:epsilonansatz}) into (\ref{genforeps}). 
Collecting orders we find

\begin{eqnarray}
\label{eqforCCC}
&&\left[-\frac{4}{5}y_0C_1+2y_1\right]\alpha g^{1/5}+\left[\left(-\frac{4}{5}C_2+\frac{2}{25}C_1^2\right)y_0+y_1C_1\right]\alpha^2g^{2/5} + \\
  &&\left[\left(-\frac{4}{5}C_3+\frac{4}{25}C_1C_2-\frac{4}{125}C_1^3\right)y_0+y_1C_2-\frac{6}{5}y_2C_1+2y_3\right]\alpha^3g^{3/5} + \nonumber\\
  &&\left\{\left[-\frac{4}{5}C_4+\frac{2}{25}\left(C_2^2+2C_1C_3\right)-\frac{12}{125}C_1^2C_2+\frac{11}{625}C_1^4\right]y_0 
+y_1C_3+\left(-\frac{6}{5}C_2-\frac{3}{25}C_1^2\right)y_2+\frac{7}{5}y_3C_1\right\}\alpha^4g^{4/5} =0 \,.\nonumber
\end{eqnarray}

\begin{multicols}{2}
The last Eq. (\ref{eqforCCC}) yields the final result for the coefficients:

\begin{eqnarray}
  C_1&=&\frac{5}{2}\frac{y_1}{y_2}~~~~~~~~=-1.153449 \\
  C_2&=&\frac{15}{4}\left(\frac{y_1}{y_2}\right)^2=\frac{3}{5}C_1^2=0.798267 
\nonumber \\
  C_3&=&\frac{95}{16}\left(\frac{y_1}{y_2}\right)^3-\frac{15}{4}\frac{y_2y_1}{y_0^2}+\frac{5}{2}\frac{y_3}{y_0}= -0.483582 \nonumber \\
  C_4&=&\frac{315}{32}\left(\frac{y_1}{y_2}\right)^4-\frac{105}{8}\frac{y_2y_1^2}{y_0^3}+\frac{35}{4}\frac{y_3y_1}{y_0^2}= 0.285279 \nonumber 
\end{eqnarray}
Using $(g^*)^{-1/5}=C_1 \alpha$, we obtain for coefficients 
$a_k$ in expansion (\ref{eqforeps2x}):
\begin{eqnarray}
a_1&=&1\\
a_2&=&C_2/C_1^2=3/5\\
a_3&=&C_3/C_1^3=0.315119\\
a_4&=&C_4/C_1^4=0.161167\,.   
\end{eqnarray}

Note that although the general method given in Appendix allows to evaluate up 
to a desired precision  {\em all}, in principle, coefficients $C_k$, it 
does not provide the closed form expression for $C_1$ as the simple approximate  
approach given in the main text does.

\end{multicols}

\begin{references}
\bibitem{remark1}
  As usual for  collision (e.g. \cite{C3,LandauLifshits})
  the two-particle problem is reduced to  scattering problem of 
  a single-particle with an effective mass $m^{\rm eff}$.
\bibitem{remark2} 
  Derivation of the dissipative force given in \cite{BSHP1,BSHP2} for colliding 
  spheres may be straightforwardly generalized to obtain the relation 
  (\ref{eldiss}) (or (A17) in \cite{BSHP1,BSHP2}) for colliding bodies of any 
  shape, provided that displacement field in the bulk of the material 
  of bodies in contact is a one-valued function of the compression 
  (see also \cite{BrilPoshprep}).
\bibitem{Tabcom}
  Obviously, the coefficients of the Pad\'e-approximation may be chosen 
  up to an arbitrary factor to multiply numerator and denominator; we chose 
  it to have unity as a leading term for both of these.
\bibitem{AttentionScales}
  Note that, in difference to the calculations in the main part of the article the quantities $x$, $x^{\prime}$ and $x^{\prime\prime}$ do have units, namely $(m/sec)^{4/5}$. The rescaled time is dimensionless. The purpose of this scaling was only to simplify the dependence of the problem on the material parameters, it was necessary to keep the explicit dependence of the problem on the initial velocity.
\end{references}
\end{document}